# MODEL ORIENTED SCHEDULING ALGORITHM FOR THE HARDWARE-IN-THE-LOOP SIMULATION

Anas M. Al-Oraiqat [*1], Yuriy O. Ivanov [2], Aladdein M. Amro[3]

[*1]Taibah University, Department of Computer Sciences and Information,
Kingdom of Saudi Arabia
E-mail: anas_oraiqat@hotmail.com
[2] Donetsk National Technical University, Ukraine
E-mail: yuriy.o.ivanov@gmail.com
[3]Taibah University, Department of computer Engineering,
Kingdom of Saudi Arabia
E-mail: amroru@hotmail.com

**Abstract:** This paper presents an approach for designing software for dynamical systems simulation. An algorithm is proposed to obtain a schedule for calculating each phase variable of a stiff system of differential equations. The problem is classified as a fixed-priority preemptive scheduling of periodic tasks. The Branch-and-Bound algorithm is modified to minimize the defined utilization function and to optimize the scheduling process for a numerical solver. A program for the experimental schedule is implemented solving a job-shop problem that proved the effectiveness of the proposed algorithm.

## I. INTRODUCTION

Computer simulation is one of the most effective methods for the study of the properties of real dynamic objects. The model complexity increases substantially for the design and research of hardware-in-the-loop simulation (HIL) of dynamic objects management systems.

Real-time simulators are classified to three different categories of applications: hardware-in-the-loop, software-in-the-loop, rapid control prototyping [1]. Among them, an important type of application is HIL [2-4]. It suggests that the real model control device be connected by I/O interfaces to manage the control object, which is implemented as an analog or digital real-time device, instead of a physical prototype. The HIL models are used in today's nuclear energy, aerospace, and defense technologies and other real-time processing [5,6].

A significant feature of the simulated processes is that their mathematical description contains a system of ordinary differential equations (ODE). These simulation tasks keep variables that must be reproduced with different frequency accuracy characteristics. The accuracy and the amount of computation depends on the chosen numerical integration methods.

The basic criteria to be met by means of simulation:
1. The frequency of parameter is changed and delivery of results must conform to the actual model [1]. At the same time, the guaranteed recovery of the simulated signals in HIL system is achieved by selecting the sampling frequency based on the Kotelnikov-Nyquist-Shannon theorem [7,8]. Frequency change of phase model variables defines the step of integration equations. In practice, for a simulation with a specified accuracy, a frequency value should be selected 10 times larger than the smallest time constant model [9,10]. In real problems, the frequency is changed in the range of hundreds of Hertz for a slow processes and hundreds of kilohertz in ultra fast transients, for example in electronics [9,10]. Such tasks require implementation of models using a step from 1 ms to 100 ms. In modern implemetations of HIL, such a step is possible and is used in electrical equipment and power systems models [11,12];
2. The simulation computational complexity corresponds to the specified performance of computing platform.
3. The periodicity of the input information leads to its cyclic processing.
4. For implementation of the model in real-time and accelerated modes [13-15], we can define $Mt$ time scale as follows:
$$Mt = Treal/Tmod \geq 1, \qquad (1)$$
where $Treal$ denotes real object or system execution time, and $Tmod$ denotes computing time for implementing the object on the modeling system.

If the performance of the simulated system is not computationally sufficient, it operates slower than real time and some real-time systems may result in catastrophic failure. The simulation feature of stiff ODE systems [16,17] is the significant difference in the rate of phase variables change that describes the object of simulation. So it is necessary to create special algorithms for real-time models simulation [18-20].

Consequently, new approaches development related to the optimization of the complex dynamic systems simulation process in real-time is citically important. It requires taking into account the implementation of new features. To create efficient models, we propose to use cyclic scheduling, which will reduce $Tmod$ for parallel computing. Hence, the aim of this research is to develop a method that optimizes software for HIL systems based on the issues raised above.

The rest of this paper is organized as follows: Firstly, Section 2 presents a justification model of the computing system organization. Secondly, Section 3 introduces the proposed computational process model. Then, Section 4 details determination of the schedule parameters. Section 5 introduces the proposed method to optimize the RT cycle followed by Section 6 that describes the implementation of the proposed algorithm to optimize the RT cycle. Then, Section 7 presents the proposed optimization algorithm. Section 8 analyzes the results of the proposed algorithm. Finally, Section 9 concludes the presented research.

## II. JUSTIFICATION MODEL OF COMPUTING SYSTEM ORGANIZATION

The digital part of model of a stiff dynamic system can be considered as a special case of a real-time task that is represented by a set of parallel threads (subtasks) calculating the phase variables in the digital part. Splitting the original algorithm of a mathematical model aims to achieve parallelism like the case of electronic analog computers. Model of Synchronous Data Flow (SDF) is often used for the formal description of the model in real-time systems [21]. The description of the process model simulation is expedient to formalize using methods of scheduling theory [22,23].

Temporary models of dynamic systems can be considered as periodic tasks that are defined by a set $(\tau i, T_i)$ where $\tau i$ and $T_i$ denote worst case of execution time and just in time (deadline), respectively. Each task must be running and fully carrying out its work in the time $\tau i$ for each respective period $T_i$ of this task. The value $\sum_{i=1}^{M} \frac{\tau_i}{T_i}$ is the CPU utilization of the set of M tasks.

The system of real-time simulation has processing threads of the object modeling, which is cyclically repeated during $Tmod$. Consider the modification of threads are already included in the model algorithm, one can select time intervals in which the simulation system must perform the same sequence of the following actions:
- Reading the input signals and generating the outputs.
- Solving ODEs' threads, taking into account the logic model.
- Exchanging results with other threads of simulation tasks.
- Waiting for the next cycle.

To synchronize data we believe that moments of time to read the input signals for all threads of the phase variables simulation are determined by the peculiarities of the dynamic model or performed into the fast thread. The interval that periodically repeats the above simulation operation is $T_c$ (the shortest repeating real-time model cycle).

Studies of periodical task scheduling allowed classification of the scheduling types and their use for various tasks that are the basis for developing hardware and software algorithms [24-27]. The basic types of periodic scheduling are [28]:
- Static Cyclic Scheduling (SCS): Its main advantages include
  - deterministic,
  - with shortest repeating cycle = least common multiple of $T_i$,
  - with a possibility to construct a static schedule within the cycle,
  - with capability of scheduling task instances according to the time-table within each cycle,
  - easy to implement.

As for its shortcoming, it is the difficulty to modify (e.g adding another task) and to handle external events.
- Earliest Deadline First (EDF): It is used for set of independent periodic tasks. Its main advantages include
  - whenever a new task arrives, sort the ready queue so that the task closest to the end of its period assigned the highest priority
  - preempt the running task
  - theoritically simple algorithm

As for its shortcomings, it is difficult to implement with the overhead of the scheduling algorithm and not predictable if any task instance fails to meet its deadline.
- Rate Monotonic Scheduling (RMS): RMS is easy to implement. Tasks are independent and always released at the start of their periods. RMS can use any fixed-priority scheduling algorithm. Usually, tasks with smaller periods get higher priorities.
- Deadline Monotonic Scheduling (DMS): One may consider DMS similar to RMS or RMS as a special case of DMS. However, in DMS, tasks with shorter deadline are assigned higher priorities.
- Handling context switch overhands: Interrupt handler runs with high priority and may delay tasks with lower priorities. The added extra time due to the system interrupts affects a system time slice and should be minimized.

From the classification of [28], the periodic schedules are divided into static, when information about the restrictions of all jobs (policy intervals) assigned to perform is known in advance, and dynamic, when jobs can be assigned to the implementation of the system during operation. Therefore important model dynamic systems that are developed and used for a long time, could reasonably be attributed to the first class schedule. In such models, the scheduler distributes the processor time among the subtasks in advance for a specific scheme. The schedule allocates subtasks to be solved in time so that they are guaranteed to satisfy all the time constraints. At the same time the scheduling process is not time critical because the schedule is generated on a preparation stage before the simulation.

In the classical theory of schedules planning [22], the start time of the periodic task is not associated with a specific point in time within the period and can vary. However, the development of simulation systems needs to ensure starting a periodic stream carried through strictly certain times periodically. That means the start time must coincide with the availability of time. The periods of tasks simultaneously start at 0.

For real problems scheduling theory does not consider scheduling periodic tasks so that each task continuously works on each of its period. Therefore, we assume that any parallel threads can be interrupted by higher priority thread at any given time on the simulation step [27]. The completion of the periodic thread during the current period $T_c$ is not interruptable. At the beginning of the next period, a new start is made, not resuming the thread. It is assumed that the overhead cost for the processor switches between threads is already included in the duration of threads. The scheduler can be implemented as a dedicated application or a thread with the appropriate (high enough) priority.

There are situations where the period or rather the release time may 'jitter' or change a little. The mentioned tasks are released at a constant rate (at the start of a constant period).

An important result of [23,29], in the theory of static scheduling for real-time systems with preemption, is to separate algorithms into two classes - algorithms with interruptions (static and dynamic priority) and control algorithms using the timeline-driven dispatching. For the algorithm of dynamic priorities, the priority of each operation may be changed when it is performed. The dynamic priorities class corresponds to the EDF family of algorithms. The basic idea of static priority RMS algorithms is that all jobs are assigned immutable priorities, which are calculated based on the known characteristics of jobs. For single-processor systems, it is proved that if a directive interval of each task is equal to the period of the RMS then the algorithm determining feasible schedule is accurate.

Most of the algorithms developed today are RMS and EDF improvements and modifications. In [30], analysis of the EDF is made and the Proportional Fair Scheduling (PFS) is proposed. The purpose of the PFS is to assign for each thread a time slice proportional to the requirements of the corresponding calculation task. Each thread call has its performance limit. The work of Anderson [31] proposed a PFS improvement, based on features of the PFS algorithm: the threads execution of each task on the entire range at a constant frequency. This behavior is achieved by splitting threads into blocks. Each block should be performed within the selected time window. The last of the intervals is the deadline for the thread model. These windows divide each part of the thread into subintervaly approximately equal length. This approach is called "Easy Release" (ER) planning. The disadvantage of the use of the ER-approach is the fact that it lost the dependence on processes speeds in the test object. Ideas of priorities and use of a proportional execution of threads at the DE system simulation step are used in this research.

To combine the strengths of the PFS and ER algorithms in [32,33], the PD2 algorithm is proposed. It sets the priorities of sub-blocks based on their deadlines. This algorithm is the best known algorithm for optimal preparation of cyclic schedules. The special features of PD2 are that the algorithm is dynamic scheduling and it concedes pre-built schedules.

The considered periodic schedules are separated subclass of cyclic tasks. Cyclic tasks are identified by the presence of closed loops in the task graph [34]. In [35], a modification of EDF algorithm for jobs with dependencies on the data (for a graph with contours) is proposed. This method does not allow to use interrupts task by higher priority threads.

The decisive factor in the construction of the schedule is the schedulability sufficient condition that depends on the time of executing the processing threads and frequency of their arrival in system. In [35,36], it is proved that the schedulability and quality of a schedule depend on the execution time of threads and frequency of the model variables. The authors propose an algorithm for constructing a scheduability cyclic schedule with a fixed priority for a single-processor computer. However, the reviewed studies do not suggest an approach to find the optimal parameters for the scheduling of the developed models. Moreover, the task scheduling optimization algorithm for the distribution of a discrete set of resources was first successfully considered by Barua [37].

When implementing the model of a dynamic system, a large class of algorithms based on SDF list-scheduling parallel jobs must be taken into account. Jobs are placed in a sorted list from which they are extracted successively and executed by a free processor. The size of the list affects the used memory size. Practically, the specific formulations of the problem and low computational complexity of algorithms need to be considered [37].

In this research, based on the description above, ideas of different algorithms classes are employed to construct an algorithm for solving the problem of the timetable for the model of the hard real-time systems.

## III. PROPOSED COMPUTATIONAL PROCESS MODEL

For the model of a dynamic system, a scheduling algorithm is considered using timeline-driven dispatching. Assume that the timeline, destination and the time of each job are known in advance.

Solution of the model development problem is performed in accordance with the detailed hardware and software of a computer system. On an abstract level, the hardware resource limitations of the model are considered and the possibility of developing the system with the restrictions implementation is analyzed.

A schedule in which all processes are carried out under specific restrictions is practically applicable. Hard real-time system limitations for cyclic process, in the absence of restrictions on memory of processors, include the following:
- Tasks are performed in the model with taking into account the parameters of objects. Therefore the mathematical description of the physical processes are represented as models of parallel jobs. The execution into threads is performed according to the selected numerical method with the appropriate step.

- A virtual processor that can run M threads is considered. Threads cyclically are called at regular intervals taking into account the worst execution time.
- All tasks must be completed before the onset of their next iteration;
- Initializing or performing other threads of the model are not required for a specific task at $T_c$.
- Each resource can be allocated to one thread implementation on a single base time interval.
- The schedule must be static with its content calculated beforehand.
- Processes can be interrupted by the timer severl times. Number of switching between processes should be minimal.
- Dependent processes and overheads are taken into account in the switching times of threads execution. All timings have integer values.

RT cycle interval is considered as a simulation step in terms of the modeling process. Each thread model includes a calculation program of one or more phase variables of the ODE system. Periods of execution threads that are corresponded to the frequencies of the phase change variables in the object may differ significantly. It determines the assignment of thread priority in the job system. Choice of cyclic scheduling discipline establishes the requirement that no one thread does not have priority relative to the other and it is provided by the order. Hence, although thread priorities are not assigned explicitly, their execution order is strictly defined. This research proposes to define it on the basis of decision-making threads deadlines, which are the values of the required periods of the simulated system. This approach of calculating state variables is consistent with the principle of EDF: the implementation of the first thread with the highest repetition rate.

*A. Model splitting*

In studying schedules for real-time simulation systems, it makes sense to consider an arbitrary interval RT cycle, as the timing of processing threads are unchanged for all RT cycles (Figure 1).

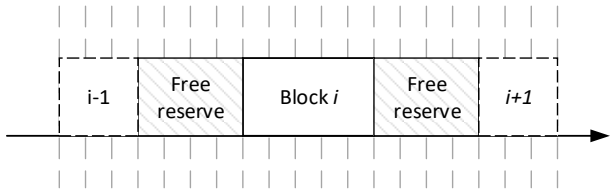

Figure 1 – Scheme of model blocks on time axis $T_c$

To organize the calculations on the basis of EDF, we propose to allocate all threads in the intervals (windows) to be performed in one RT cycle. This is due to the fact that, when processing thread of high frequency with slow threads, a situation will necessarily arise with the execution of the slow thread violating the deadline for performing fast. In this case, a slow thread can be decomposed into blocks. Each block is processed in the next provided window $T_c$. Thus each of the thread model is represented by a set of composite blocks on the preparation phase. Each block contains an iteration part of the simulation task. The slice of a part depends on two values: RT cycle timings and task deadline $T_i$. Each block begins and ends with the context switch interrupt of another block from another thread.

A developer of a dynamic object model performs software implementation of a task threads and generates a schedule that determines the timings for RT cycle blocks. The only constraint for each thread is the RT cycle boundary $T_c$.

Threads management routine according to the generated schedule is implemented for the suitable Operating System (OS). The required set of OS functions are: running thread, pausing and resuming execution, communicating between threads, and working with timer. Context switch overheads for all threads are considered constant and should be determined for each specific modeling system. These time delays are caused not only by the OS functions, but also by hardware implementations of algorithms, read/write memory, context switching, and cache misses on the data. Analytical determination of the delay is difficult for the software part (as most OS are closed) and for hardware (as processors' developers do not provide structural schemes and algorithms of their devices). Determination of the numerical values of the delays can be done experimentally with the help of specially developed profiling performance test.

The executive part of the simulation system has the following initial conditions:
The CPU calculates the time sequence diagram of the M threads for execution, each of which is the necessary CPU time $\tau_i$ $(i = 1,2, ..., M)$ with period $T_i$ $(i = 1,2, ..., M)$. Each thread must be executed until the next $T_i$. The necessary condition for the existence of schedules is as follows [16,32]:

$$\sum_{i=1}^{M} \frac{\tau_i}{T_i} \leq 1 \qquad (2)$$

For stiff ODE systems [21], the value of thread processing periods can be sorted by ascending order: $T_1 < T_2 < ... < T_M$. Let $L = \text{GCD}(T_i)$ (greatest common divisor), base period RT cycle, during which a specific part $\Delta i$ $(0 < \Delta i \leq 1)$ of each thread is performed. Each thread performance is synchronized by $L$. The frame for the execution of each block is defined as $\Delta_i \cdot \tau_i$. Thread will be executed fully for $k_i = ]\frac{T_i}{L}[$ RT cycles, where $k_i$ is defined as the largest integer less than or equal to $k_i$. With this organization, the RT cycle is actually replacement periods $T_i$ threads in the system to values $T_i' = k_i \cdot L \leq T_i$, as multiples of L. Additionally, decrease of $T_i$ is explained by the fact that, there are situations, where the period or rather the release time may 'jitter' or change a little, but the jitter is bounded with some constant. The "jitter" may cause some tasks missing deadlines. So it is possible to manipulate the periods so that they are multiples of each other. Then the cycle will be complete $k = ]\frac{T_c}{L}[$ RT cycles. In this case, it is easier to find a feasible schedule and reduce the size of the static schedule with less memory usage. The described cyclic schedule can be represented as a list of block execution in the timing diagram of a full period $T_c$ as shown in Figure 2.

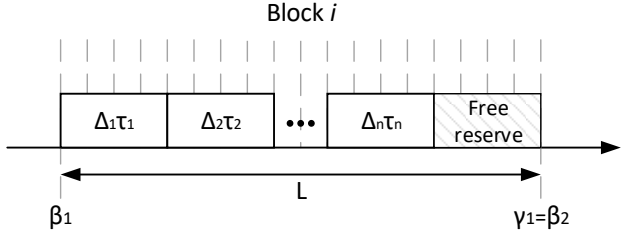

Figure 2 - Scheme of the computational process organization for one full period $T_c$

In accordance with the definitions [36], where $α_j$ is the time of a request for the thread execution of $j^{th}$ cycle, $β_j$ is the processing beginning of the $j^{th}$ RT cycle, and $γ_j$ is the end execution time of the $j^{th}$ cycle, the existence conditions of a RT cycle are:
1. Blocks of thread in each RT cycle are the same:
$(Δ_iτ_i)_j = (Δ_iτ_i)_{j+r}$ $(i = 1,2, … n; j = 1,2 …; r = 1,2, … k)$ (3)

2. The cyclic schedule does not consider arrival of requests: $β_j = α_j (j = 1,2, … , k)$.

3. Processing of the entire thread group is executed on every base cycle:
$β_{j+r} = β_j + rL$ $(r = 1,2, … k)$.

4. Processing more than one application simultaneously can not be executed.

5. RT cycle execution must be completed no later than the moment of arrival of the next group of threads: $T_c − γ_k ≥ 0$.

A cyclic schedule is admissible, if the conditions listed above 1 to 5 are fulfilled. However, block sizes of threads are reduced. The proposed organization enables flexible modification of the model and the implementation of background tasks to free intervals of $T_c$. Free intervals can be used for the model in an accelerated time scale. In general, the correct schedule is not unique to the system. Scheduling is usually determined by the extremum of the objective function.

## IV. DETERMINATION OF THE SCHEDULE PARAMETERS

To develop an effective schedule, it is necessary to determine the optimal parameters of the RT cycle, particularly the value of the base period of the cycle $L$. The criterion for the efficiency of the system is its workload. In comparison with an ideal schedule, the proposed efficiency criterion considers the additional costs of CPU time for the organization RT cycle. This utilization growth (μ) is the performance difference in the threads timings on the period of the RT cycle:

$$μ = \sum_{i=1}^{M}(\frac{τ_i}{T'_i} − \frac{τ_i}{T_i}) \quad (4)$$

Considering time costs in this expression, it is necessary to replace $n$ by $\frac{np}{L}$. Thus, the efficiency of the schedule can be estimated as the following function:

$$F = \sum_{i=1}^{M}[(\frac{τ_i}{T'_i} − \frac{τ_i}{T_i}) + \frac{np}{L}, \quad (5)$$

where $p$ is the average overhead of switching one thread and $n$ is the number of threads.

A sufficient condition for the existence of schedules and the restriction on the change in the parameter base period $L1 ≤ L ≤ T_1$ of RT cycle is used for $F$. The value of $L$ under the terms of the technical feasibility is a natural number. A formal description of the model to optimize the computational process can be written as:

$$\arg\min_{L=1,2,…,T_1} F(L) \quad (6)$$

$$F = \sum_{i=1}^{M}[(\frac{τ_i}{T'_i} − \frac{τ_i}{T_i})] + \frac{Mp}{L}$$

$$\sum_{i=1}^{M}\frac{τ_i}{T'_i} ≤ 1, \ T'_i = ]\frac{T_i}{L}[L, i ∈ ℕ, i ≤ M$$

The optimization model of (6) can be used for calculating the numerical values of the model parameters. A feature of the model objective function of (6) is that it includes two components:

$$F_1 = \sum_{i=1}^{M}\left(\frac{τ_i}{T'_i} − \frac{τ_i}{T_i}\right), \ F_2 = \frac{np}{L} \quad (7)$$

The first part ($F_1$) determines the increase of the CPU utilization due to changes in the base thread periods. Based on $T'_i$ values, the function $F_1$ is nonlinear. The minimum allowable value of $F_1$ is obtained when the period of performance of the threads does not change, i.e., if $T'_i = T_i$, $\min(F_1) = 0$.

$F_2$ function is hyperbole, that is increasing L results in reduction in the CPU time cost of switching between threads. Restriction for the values of $p$ is determined from the relation:
$$F_2(L = T_1) = 1 − F_1(L = T_1) \quad (8)$$

Optimized parameters of the RT cycle model are transferred to the scheduler of the HIL model as time sequence diagrams. At this point, the training phase of the algorithm is completed. Hence, the model becomes ready to run.

## V. DEVELOPMENT OF A METHOD TO OPTIMIZE THE RT CYCLE

Analysis of the mathematical model to optimize the schedule shows that it belongs to a class of integer nonlinear programming problems [38,39]. This is based on the fact that the function $F$ is nonlinear and multiextremal. The total number of satisfying solutions for the problem is the Stirling number of the input sequence $T_i$. This problem is NP-complete. To solve this problem, universal techniques of nonlinear programming can be used, but they can only find local extrema. For extrema search of common tasks, decomposition to simpler subtasks with linear complexity can be used.

Considering features of HIL, the optimal value of $L$ is the largest value not exceeding $T_1$ for the RT-cycle, which is consistent with the necessary condition for the existence of schedules. Accordingly, the initial value and search direction can be set.

Strict limitations on the existence of the schedule may result in that a schedule for a given value of $L$ can not be constructed. In this case, in the process of optimizing, the $L$ further continuation no longer makes sense, and hence this option can be discarded. This approach is known as the technique of "sequential analysis, design and filtering out variants". In this method of variants construction, unpromising solutions are eliminated without their full completion.

Based on the method of Branches and Borders (B&B) [39], efficient algorithms are to be developed for solving the problem using known optimization algorithms. Generally, B&B is a tree-based optimization method that uses four operations (selection, branching, bounding and pruning) to build and explore a highly irregular tree representing the solution space [40]. The B&B method guarantees finding the exact solution of the problem and allows taking into account the additional restrictions on the schedule.

Accounting for the differences between the mentioned methods, a new algorithm based on the B&B method is developed. The main problem of the algorithm is to develop criteria for evaluating the upper and lower bounds for the solution optimal values for subregions of the search tree.

An optimization feature of the problem is that the two parts of the objective function of (6) vary nonlinearly. However, changing the first part of $F$ is not associated with a change in its second part. Consequently, the decision can be made at the decomposition of the set of admissible search plans for optimal solutions to subsets. This is to be done with consistent calculation of objective functions estimates for each subset. The intermediate values of the objective function are specified in the following calculations. Marked limitations of the lower and upper values of the objective function allow to cut off those values in the solutions that do not correspond to the problem constraints and can not be considered in the future.

Let $G$ be a finite set of solutions of the objective function of (6)

$$F = f(L), \ L \in G. \qquad (9)$$

The general proposed scheme for solving the problem of B&B method is described through the following cyclic sequence of steps:
1. Calculate the lower limits of the objective function $f(L)$ on $G$ and its subsets.
2. Split the set G to subsets of tree.
3. Calculate the lower limit of $f(L)$ on subsets.
4. Calculate admissible plans.
5. Check for the optimal plan.

## VI. ALGORITHM IMPLEMENTATION TO OPTIMIZE THE RT CYCLE

Consider an algorithm for solving the problem, subject to the limitations and peculiarities. Let there be a set $H = \{H_{g1}, H_{g2},...,H_{gT_1}\}$ that contains all the elements of a subset of the objective function $\sum_{i=1}^{n}(\frac{\tau_i}{T_i^{'}} - \frac{\tau_i}{T_i})$ for all admissible planned changes of $L \in [1,T_1]$. Each subset can be represented as $H_{gj} = \{\frac{\tau_1}{T_1^{'}} - \frac{\tau_1}{T_1}, \frac{\tau_2}{T_2^{'}} - \frac{\tau_2}{T_2},..., \frac{\tau_n}{T_n^{'}} - \frac{\tau_n}{T_n}\}$. Let then the set $V$ contains all $\frac{np}{L}$ elements for planned changes of the variable $L$. The first element of the set is equal to $\frac{np}{L}$ with $L = T_1$. All other elements are the modulus of the difference relative to the previous element of $V$. Thus, $V = \{\frac{np}{T_1}, V_{\Delta}^{T_1,T_1-1},...,V_{\Delta}^{2,1}\}$ is the set of feasible solution. The expression $V_{\Delta}^{i,i-1} = \left|\frac{np}{L} - \frac{np}{L-1}\right| = np\frac{1}{L(L-1)}$ is valid for the variable $L \in [1,T_1-1]$. The upper and lower estimates of the objective function can graphically be represented as follows (Figure 3):

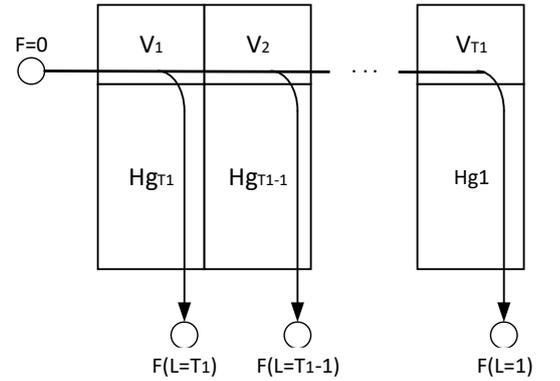

Figure 3 - Diagram for determining the objective function $F$.

The shown tree of Figure 3 has a root at $F = 0$. The branches of the tree are the elements of the $H$ and $V$ sets. From the point $F = 0$, the value of the objective function is determined by summing up all the subsets of elements along the lines of the arrows. After each addition, the objective function constraints should be checked. If they are exceeded, the considered solution is no more feasible and has to be rejected. Assume that set C includes all elements that were summed in F before the current addition step. Initially, the set $C$ is empty ($C = 0$). Next, the element $V_1$ of the set $V$ is entered into $C$ leadind to $C = \{V_1\}$. This element is always taken into account for all possible values of the variable $L$, even if no feasible solution has not been selected yet. The estimate of $F$ will include only element $V_1$ that is $F(V_1)$ (Figure 4).

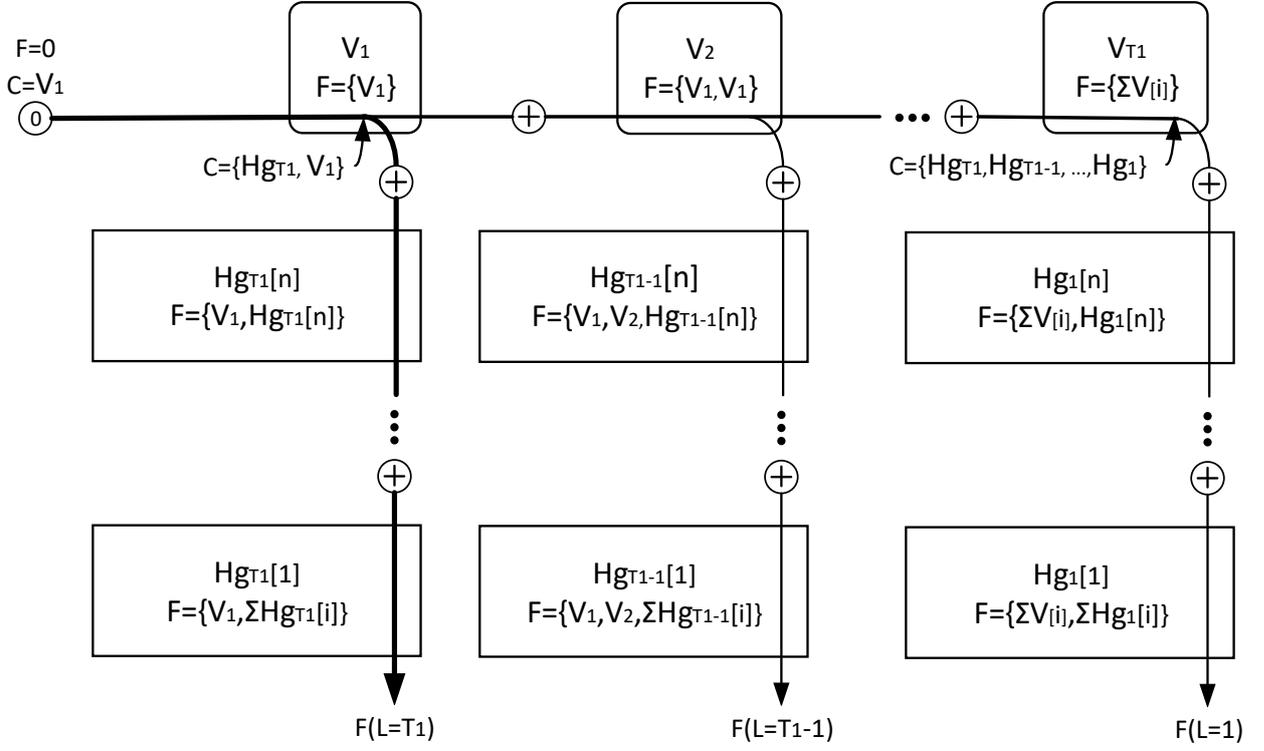

Figure 4 - Search Decision Tree

The next step is the first branching that corresponds to a split decision into two variants:
1) $L = T_1$ (down the tree) corresponding to a subset $H_{gT_1}$.
2) $L \in [1, T_1 - 1]$ (right) for all other variants corresponding to the remaining elements of the $H$ and $V$ sets, as well as the root element $V_2$ of the remaining subtree solutions.

Thus, the contents of the set $C$ changes to be $C = \{H_{gT_1}, V_2\}$. After that, the estimate in the current step (compared to the value of the objective function for $F(V_1, V_2)$ and $F(V_1, H_{gT_1}[M])$) is determined. For the calculated values of the subset $H_{gT_1}$, the last element is always selected. This search direction is due to having the elements of $H$ corresponding to non-linear ascending function of module division. Thus, non-optimal feasible solutions can be weeded out at the initial steps of the solution. In this case there are two possibilities:

1) If $F(V_1, V_2) \geq F(V_1, H_{gT_1}[M])$, then the choice of the subset variant $H_{gT_1}$ with $L = T_1$, is optimal. In this case, consideration of the set $H_{gT_1}$ continues with procedure of unilateral branch. The set for partitioning is selected among final subsets of the previous step of branching, for which the value of $C$ does not change: $C = \{H_{gT_1}, V_2\}$. The next step considers the next element of subset $H_{gT_1}[n-1]$, $F(V_1, V_2)$ and $F(V_1, H_{gT_1}[n-1])$.

2) If $F(V_1, V_2) < F(V_1, H_{gT_1}[n])$, then the current content of $C$ determines that the choice of solution $L = T_1$ is not optimal solution. Therefore, it is necessary to consider a range of variants $L = [1, T_1 - 1]$. In this case, the procedure of branching and the set $C$ contains $C = \{H_{gT_1}, H_{gT_1-1}, V_3\}$ with new branches $H_{gT_1-1}$ and $V_3$ instead of $V_2$. That means that new feasible solution was added that considers solution fot thr value $L = T_1 - 1$ and solutions for the rest of the subtree, consider $L = [1, T_1 - 2]$.

The cyclic sequence of such actions will result in an optimal value for the objective function. It is reached when the first element of any subset $H_{gi}[1]$ is selected in C. If the value $V_{T_1}$ reached, but the solution is still not found then the algorithm proceeds exclusively in one-way branching for all subsets until obtaining an optimal value of the objective function.

In a refinement step of the objective function $F$, the maximum limit of the objective function may be achieved. In this case, calculation for this branch is terminated and not considered further. The maximum lower boundary of (6) is $F = 1$. So, the solution, for which $F \geq 1$, is infeasible and it can not provide needed schedule. Therefore, the corresponding branch and set $H_{gi}$ are removed from the tree.

The proposed algorithm searches for the minimum of the objective function by dividing the search task to more simple tasks. This division makes it possible to search for the solutions step by step. At each step, it is possible to perform pruning using the B&B scheme.

## VII. OPTIMIZATION ALGORITHM

The following algorithm is proposed for finding the minimum of the objective function. $valF[T_1]$ is a vector that determines the current value of the objective function for all considered candidate solutions. Sets *setC, setMinC* and *setX* contain subsets of elements $V$ and $H_g$ required for the analysis of each iteration of the algorithm. *setC* defines a set of elements to be considered on the step of the algorithm. *setMinC* includes the minimum value from *setC* that was selected on the current iteration of the algorithm. *setX* contains a set of child elements connected to the edge currently included to *setMinC* according to the structure (Figure 3).

Consider the following sequence of steps of the algorithm:

Assume that function *FindSet(Z)* returns all child element connected to Z. The first step is to find the *setX* that means elements connected to *setMinC* or *FindSet(setMinC)*. If $setMinC = \{\}$ then $setX = \{V_1\}$ is chosen as the root. The *setMinC* is excluded from *setC* and *setX* is included to *setC* because now these values should be considered in the next step. Also the set is checked for being empty or not? This case is possible, when all the branches have been checked, but the valid value has not been found. Then after applying *FindSet* to *setX*, *setX* will be {}, because none of the subsets $V$ and $H_g$ remains unchecked. The next element from set *setC* is excluded, considered the last step *setMinC*. The set *setC* after merging with *setX* is empty. In this case, the algorithm is terminated because finding the solution of the problem (6), for the given set of input data, is impossible.

The second step is to find the minimum element in the *setC*. The value of this element is included to the *setMinC*. Determination of minimum is performed by comparing the values of all elements of the set.

In the third step, the algotithm continues to calculate the value of the objective function for the selected possible solution, by adding this minimum value of *setMinC*.

In the fourth step, an elimination test is performed. It checks whether a value exceeds the maximum limit of the objective function or not? In this case, the current branch is excluded from future considering.

In the fifth step, the branch contained in *setMinC* is checked if it is the first element of any feasible solution $H_{gi}[1]$, the calculation stops and it means that $i^{th}$ feasible solution is optimal (*L=i*), otherwise jumps to Step 1.

The algorithm can be represented by a graph diagram of Figure 5.

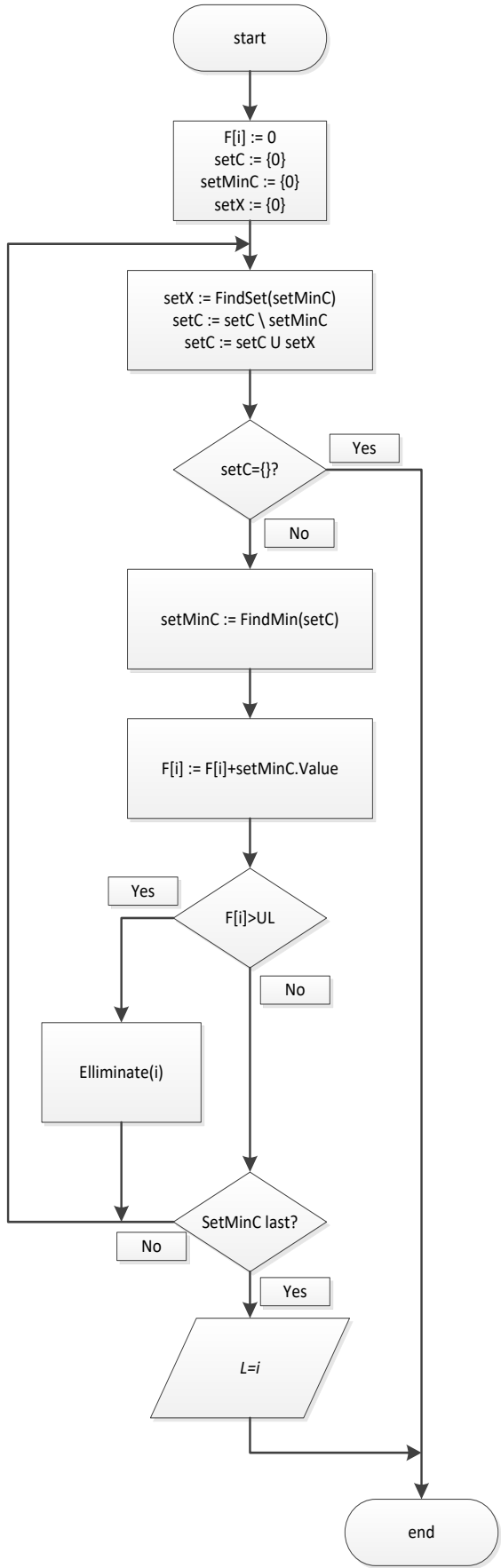

Figure 5 - Algorithm to find the optimal solution

The implemented algorithm performs search for the optimal solutions of the solution and determines the desired value of L used to generate the schedule.

## VIII. ANALYSIS OF THE RESULTS.

The computational complexity of each individual function implemented in the algorithm depends on the number of threads (M) and the period of the fast thread ($T_1$). The fastest changing function is either *FindSet* or *FindMin* depending on which of the values $(M, T_1)$ is greater. Since the optimization problem is non-linear and its solution scheme uses B&B, then its evaluation of computational complexity will be the upper boundary or all the possible options. The complexity of our implemention on C# for finding the base period RT cycle is defined as $O(\min(M, T_1) * M * T_1)$.

The proposed algorithm is optimum in terms of memory requirements. In the worst case all sets considered in our implementation will use less than $T_1 + 5$ instances of the edges represented in the structure of Figure 3 in memory.

For example of solving the problem of determining the schedule, consider the next set of input parameters (*M*=4, $\tau_i = \{1,3,3,4\}$, $T_i = \{5,16,19,22\}$, *p*=0.2). Consistent execution of the algoritm iterations provides solution *L* = 5 with the value of the objective function $F[5] = 0.2325$. The calculation of the objective function was performed for other values of *L* as well. The results of the calculations are presented in Table I.

Table I: Results of the calculations of the objective function.

| L | 5 | 4 | 3 | 2 | 1 |
|---|---|---|---|---|---|
| F | 0.232 | 0.297 | 0.429 | 0.458 | 0.800 |

To evaluate the efficiency of the algorithm, generating the input test sequences was performed. As Sezare theorem discussed for generating a series of natural numbers [41], approximately 60% of cases obtain a pair of mutually prime numbers.

The effectiveness of the proposed algorithm was evaluated as the ratio of the number of steps expended on finding L, compared with the solution of brute force. The program has been run 100 times for random input data, wherein the efficiency was on the average 38.87% higher than the case of exhaustive search [42].

To assess the effect of non-multiple periods in the algorithm, special numerical sequences have been considered. For this purpose Fibonacci number series and prime numbers have been selected. The average efficiency of the algorithm is 29.77% for a sequence of prime numbers and 44.64% for Fibonacci numbers.

## IX. CONCLUSIONS

In this paper, we have analyzed real-time schedulers and their features that can be used for numerical integration of the stiff dynamic systems. The research proposed a new time model as a set of periodic parallel tasks (threads) that calculate the phase variables of the system of ordinary differential equations. The main feature is accenting the research on a multi frequency nature of simulation systems.

We have presented a new scheduling policy that belongs to the so-called Self-Timed Periodic scheduling. This schedule improves performance, decreases synchronization costs, resource sharing and resource constraints. The schedule optimization is a combinatorial optimization problem. Its schedulability is used to define the objective function and the constraints of the system. This problem is NP-complete. To solve it, we adapted a branch-and-bound algorithm. The proposed computational scheme represents a one-way branching tree. We have implemented the proposed algorithm on the C# and verified the optimization approach. Results were confirmed by comparison to manual test sets of input parameters. Analysis of this software shows the efficiency of the algorithm.

The proposed optimization approach allows us to generate efficient schedules for the stiff dynamic systems that is used later by a process (threads) manager in real-time operating systems for HIL systems parallel simulation.

## X. ACKNOWLEDGMENT

The authors would like to thank Taibah University and Donetsk National Technical University for supporting this research.

## AUTHOR'S PROFILE

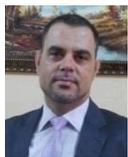 **Anas M. Al-Oraiqat** received a B.S. in Computer Engineering and M.S. in Computer Systems & Networks from Kirovograd Technology University in 2003 and 2004, respectively, and Ph.D. in Computer Systems & Components from Donetsk National Technical University (Ukraine) in 2011. He has been an Assistant Professor at the Computer & Information Sciences Dept., Taibah University (Kingdom of Saudi Arabia) since Aug. 2012. Prior to his academic career, he was a Network Manager at the Arab Bank (Jordan), 2011-2012. Also, he was a Computer Networks Trainer at Khwarizmi College (Jordan), 2005-2007.

His research interest is in the areas of computer graphics, image/video processing, 3D devices, modelling and simulation of dynamic systems, and simulation of parallel systems.

E-mail: anas_oraiqat@hotmail.com

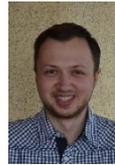 **Yuriy O. Ivanov** received a M.S. in Computer Engineering from Donetsk National Technical University (Ukraine) in 2010, and Ph.D. in Computer Systems and Components from Donetsk National Technical University in 2013. He had been an Assistant Professor at the Computer Engineering Dept., Donetsk National Technical University till Sept., 2014. Since then he has been working as a software engineer in data protection area.

His research interests include multithreading and scheduling, real-time systems, simulation of dynamic systems.

E-mail: yuriy.o.ivanov@gmail.com

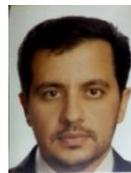 **Aladdein M. Amro** received M.S. in Automation Engineering from Moscow Technical University in1996, and Ph.D. in Telecommunications Engineering from Kazan State University (Russian Federation) in 2003. Had been an Assistant Professor at the Computer Engineering Dept. Al-Hussein Bin Talal University (Jordan) during the years 2004-2011. Since then has been working as an Assistant Professor at the Computer Engineering Dept., Taibah University (Kingdom of Saudi Arabia).

Research interest is in the areas of digital Signal processing, image processing, real time systems.

E-mail: amroru@hotmail.com